# Analytical Study on Internet Banking System

Fadhel.S.AlAbdullah, Fahad H.Alshammari, Rami Alnaqeib, Hamid A.Jalab, A.A.Zaidan, B.B.Zaidan

**Abstract** – The Internet era is a period in the information age in which communication and commerce via the Internet became a central focus for businesses, consumers, government, and the media. The Internet era also marks the convergence of the computer and communications industries and their associated services and products. Nowadays, the availability of the Internet make it widely used for everyday life. In order to led business to success, the business and specially the services should provide comfort use to its costumer. The bank system is one of the most important businesses who may use the website. The using for the web-based systems should contain special requirements to achieve the business goal. Since that the paper will present the functional and non-functional for the web-based banking system.

**Index Terms**— Internet Banking System, Analysis, Functional Requirements, Non-Functional Requirements, Functional Requirements Specification.

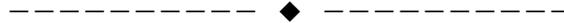

## 1. INTRODUCTION

Consumer behavior is changing partly because of more spare time. The way of use of financial services is characterized by individuality, mobility, independence of place and time, and flexibility. Financial transactions caused by purchases will more and more be carried out by non- and near-banks [1]. These facts represent big challenges for providers of financial services. More and more the Internet is considered to be a "strategic weapon". Financial services companies are using the Internet as a new distribution channel. The goals are:

- complex products may be offered in an equivalent quality with lower costs to more potential customers;
- There may be contacts from each place of earth at any time of day or night.

This means that financial institutions may enlarge their market area without building new offices or field services, respectively[2]. Because of its image as an innovative corporation, better interacting possibilities, the usage of rationalization potentials, promotion of self-serviced as, the improvement of its competitive situation by development of core competencies together with the construction of market entry barriers, it may be possible to increase profits and market shares.

One way of exploiting rationalization potentials is the implementation of the entire transaction (from purchase to payment) under a common user interface. Information collected in operative databases of financial institutions allows them to act as information brokers. Offering special information in closed user groups may result in more intense customer commitment, as well as customer bonding. Know-how that is built up by Internet presence may be used to facilitate Internet presence of smaller companies. The use of digital coin-based money to completely settle transactions in the Internet is a new service provided by financial institutions [4].

In the recent years there has been explosion of Internet-based electronic banking applications states that the emergence of new forms of technology has created highly competitive market conditions for bank providers. However, the changed market conditions demand for banks to better understanding of consumers' needs

In [5] stress that the success in Internet banking will be achieved with tailored financial products and services that fulfill customer' wants, preferences and quality expectations. concedes that customer satisfaction is a key to success in Internet banking and banks will use different media to customize products and services to fit customers' specific needs in the future. Suggest that consumer perceptions of transaction security, transaction accuracy, user friendliness, and network speed are the critical factors for success in Internet banking. From this perspective, Internet banking includes many challenges for human computer interaction (HCI).

In [6] have remarked that there are at least two major HCI challenges in Internet banking. The first challenge is related to the problem how to increase the number of services of Internet banking and simultaneously guarantee the quality of service for individual customers. The second challenge is related to the problem how to understand customer's needs, translate them into targeted content and present them in a personalized way in usable user interface [7]. Imply that Internet banking research will concentrate more on HCI factors in the future. Recently, emphasize that now is an ideal time for HCI researchers to analyse user satisfaction, because there is growing interest in how to attract and increase the number of online customers in e-business and e-commerce. Stress that HCI researchers should reveal a structure of user satisfaction, determine how to evaluate it and conclude how it is related to the overall user experience of online customers. The concept of electronic banking has been defined in many ways electronic banking is a construct that consists of several distribution channels. Defines electronic banking as the delivery of banks' information and services by banks to customers via different delivery platforms that can be used with different terminal devices such as a personal computer and a mobile phone with browser or desktop software, telephone or digital television [8],[9].

## 2. METHODOLOGY

## 2.1. Functional Requirement

General principles of requirements engineering is a distinction between requirements definitions and requirements specifications.
A software requirements definition is an abstract description of the services which the system should provide and the constraints under which the system must operate. And requirements definition is probably the most important technique in structured analysis. It is the only technique that permeates every step of the method. It also is one of the least pictorial, making it difficult to describe precisely. In essence, the technique involves capturing what the users really want and making sure that every subsequent project activity leads to the best possible transformation of those user needs into system needs which, when satisfied, will deliver what the users wanted in the first place.

• **Customer**

The valid customer on internet banking has a set of requirements he/she does on internet banking. These requirements are offered on next pointes.

• **Login**

A customer to be able to use this system, he/she has to enter username and password which he/she has created before and been saved in the database in the Login page. This function might be a customer or an Admin also.

The input in this function most be valid username and valid password and the output if the user is valid user then he/she will get into a page which can makes has/her transaction, but if the user made wrong in username or password then he/she will be invalid user and will see a message "Alert Invalid Username and Password" and to login again.

• **View Account**

View Account allows to a customer to view today's up-to the minute balance information on deposit (saving/current), credit card, etc. The customer can also view transaction history with retention period up to a maximum of 90 days. Within this feature, the customer can request for account such as "view online, by e-mail or by post option. But the customer most be logged in the internet banking.

• **Transfer Funds**

The customer must be logged into Banking System to be able to make his/her transaction for transfer funds. Transfer Funds allows customer to transfer funds between authorized accounts – own personal accounts. Requested transfer take place immediately or at a selected future date specified by customer. The customer can save up to a maximum of 10 accounts and update or delete the account details. All the outstanding future transfers are recorded in a table. The customer can enquire whether there is any funds transfer pending and. when the customer selects the Transfer funds, the system will display Menu to select Transfer Funds function for transfer funds or Transfer History function for display the transaction he/she done.

• **Pay Bills**

The customer most be logged into Banking System. With internet banking, customers can make payments to corporations that include utilities, assessments, Insurance, telecommunications, and other services. The customers can use Online Pay Bill service to pay bills by debiting their account. This payment made to payee corporations that the customer has registered with internet banking by using the Registered Bill. But with new payee corporations that the customer has not registered, this payment can be made immediately or at a later date.

The customer needs to key in his/her bill account number each time you make a payment. Also the customer makes payment (up to the outstanding balance) to his/her owns credit card and balance transfer account. And he /she can register the bills. After the registration, uses "Registered Payment" for subsequent payments by Bill registration. He /she doesn't have to enter his/her bill account number anymore. And remove bills from list of "Registered Payment" by using the Bill Deregistration function. For Pay Bill Any bill can be changed or canceled, so There are Enquiry Future Payment Status, this function lets customer enquires whether if has scheduled any future payments or not. And Cancel Future Payment lets customer cancels his/her scheduled future payments if he/she changes his/her mind.

- **Cheque Services**

The customer most be logged into Banking System.The customer may enquiries cheque status, whether it is paid, unpaid, stopped or returned. It also allows customer to stop cheque payment and to request for a cheque book online.

- **Utility**

The customer most be logged into Banking System. Utility allows customer to change password and the secure delivery contact information. Within this feature, the customer can also change the online profile personal information that is retained by the internet banking system only. And the customer can cancel the ATM facilities.

- **Logout**

The customer most be logged into Banking System. This function is used when a logged in user finishes his/her job and wants to be logged out so that no one can abuse his username. The system will state the user has been logged out successfully.

- **Administrator**

An administrator is that person who makes some editing for the internet banking system like add/cancel customer, check the transactions etc. but this administrator must be valid user. Therefore the administrator must have a username and password. In the project we will not go deep in an administrator because we will focus on the customer and his/her requirements more than the administrator.

## 2.2. Functional Requirement Specification

Requirements specifications add further information to the requirements definition. Natural language is often used to write requirements specifications. However, a natural language specification is not a particularly good basis for either a design or a contract between customer and system developer.

A Functional Requirements Specification describes what is required to meet the users' business needs. Functional requirements specify which actions the design must provide in order to benefit the system's users. Functional requirements are determined by the needs, user, and task analysis of the current system.

➢ **Customer**

We mention to what the customer needs to do on internet banking system and we are going to go through these needs and how the customer can do it.

✓ **Login**

**Definition:** For the users to be able to use this system, they have to enter username and password which they have created before and been saved in the database in the Login page. The user might be a customer or an Admin also.

**Inputs:** Username and password.

**Outputs**: The system will state whether inputs are correct or not.

**Pre conditions:** The user must have signed in the system and have a valid username and password. Then the system will show the main page to the valid customer and display message "welcome to the internet banking system please click on the left menu bar to choose your option!" he/she can make has/her transaction, but if the user made wrong in username or password then he/she will be invalid user and will see a message "Alert Invalid Username and Password" and to login again.

**Post conditions:** The user will enter the main page of him/her self.

✓ **View Account**

**Definition:** View Account allows to a customer to view today's up-to-the minute balance information on deposit (saving/current), credit card, etc. The customer can also view transaction history with retention period up to a maximum of 90 days. Within this feature, the customer can request for account such as "view online, by e-mail or by post option. But the customer most be logged in the internet banking.

Inputs: there are not inputs in this function.

**Outputs:** the system will show the View Account page and display a message" Please click on the respective account/card types for more details. Customer can choose current account or saving account for more details.

**Pre conditions:** The customer must be a valid customer and signed in the system.

**Post conditions:** The customer clicks on the logout button or select other functionality options.

✓ **Transfer Funds**

**Definition:** Transfer Funds allows customer to transfer funds between authorized accounts – own personal accounts. Requested transfer take place immediately or at a selected future date specified by customer.

**Inputs:** amount, target account and TAC. Also if he/she wants to enter his/her e-mail, and select the current account or saving account.

**Outputs:** the system will display Transfer Funds function for transfer funds or Transfer History function for display the transaction he/she done.

**Pre conditions:** The customer must be a valid customer and signed in the system.

**Post conditions:** The customer clicks on the logout button or select other functionality options.

➢ **Pay Bills**

**Definition:** The customer selects the Bill Payment functionality then the system displays Bill Payment Menu, and the customer selects one of four functionalities from Bill Payment menu.

✓ **Pay Registered Payment**

This function allows a customer to pay Immediate and future payment to corporations, those customers have registered when he/she selected it.

**Inputs:** Select Corporation Name from the list provided and enters the payment Amount (without "RM) and Bill Reference Number, if required. And Select Effective Date.

**Outputs:** the system will display the **Confirm** message and show the payment details.

✓ **Open Payment**

This function allows a customer to pay Immediate and future Payment to corporations that customer has not registered.

**Inputs:** Select Corporation Name either:
1. Select from Top Ten Payees list provided if the corporation you would like to pay is in this list.
2. The customer can write the payee name. Then enter **Account No,** and Enter your bill account holder name, bill account number and payment amount and bill reference number if required and select Effective Date.

**Outputs:** the system will display the **Confirm** message and show the payment details.

✓ **Pay Registration/Delete Registration Bills**

1. **Pay Registration Bill**

Select Corporation Name from the list provided Enter the Bill Account Number, Bill Account Holder Name. And key information required, and then click' **Register**. The system will appear the confirm message to confirm the transaction. The status and details of the customer's registration bill will appear.

2. **Pay Delete registration Bill**

When the customer selects and clicks on **Deregistration Bill**. The screen will display all the registered bills. Tick the box of payee(s) the customer intends to delete from the list and click cancel. Then the system will appear message confirmation to confirm the transaction.

✓ **Bill Payment History**

If the customer wants to display his/her payment history just he/she has to click on Bill Payment History, the system will display the transaction he/she done.

**Pre conditions:** The customer must be a valid customer and signed in the system for all these functions.

**Post conditions:** The customer clicks on the logout button or select other functionality for all these functions.

➢ **Cheque Services**

**Definition:** The customer may enquiries cheque status, whether it is paid, unpaid, stopped or returned. It also allows customer to stop cheque payment and to request for a cheque book online.

**Inputs:** cheque number.

**Outputs:** the system will display the confirm message and show the details transaction.

**Pre conditions:** The customer must be a valid customer and signed in the system.

**Post conditions:** The customer clicks on the logout button or select other functionality options.

➢ **Utility**

**Definition:** Utility allows customer to change password and the secure delivery contact information. Within this feature, the customer can also change the online profile personal information that is retained by the internet banking system only. And the customer can cancel the ATM facilities.

**Inputs:** In change password function a customer should enter new password and IC/Passport No. and in update profile the customer should change information that he/she wants to change.

**Outputs:** The system will show the user has been logged out successfully.

**Pre conditions:** The customer must be a valid customer and signed in the system.

**Post conditions:** The customer clicks on the logout button or select other functionality.

➢ **Logout**

**Definition:** This function is used when a logged in user finishes his/her job and wants to be logged out therefore, that no one can abuse his/her username.

**Inputs:** there are not inputs in this function.

**Outputs:** The system will show the user has been logged out successfully.

**Pre conditions:** The user should have logged into the system.

**Post conditions:** The user will enter the main page of the system.

## 2.3. Non-functional Requirements

Non-functional requirements are requirements that are not directly concerned with the specific functions delivered by the system. They may relate to emergent system properties such as reliability, response time and store occupancy. They may specify system performance, security, availability, and other emergent properties. This means that they are often more critical than individual functional requirements. System users can usually find ways to work around a system function that doesn't really meet their needs. However, failing to meet a non-functional requirement can mean that the whole system is unusable. Non-functional requirements needed in this internet banking system are identified as performance requirements, safety requirements, security requirements and software quality attributes.

## 2.3.1 Performance Requirements

• Increase Customer Satisfaction
Internet banking system must allows customers to access banking services 24 hours a day, 365 days a year with minimum downtime period for backup and maintenance.

• Expand Product Offerings
The new services allows bank to capture a larger percentage of their customers' asset base. The internet banking system will provide facilities for bank to offer new services and products onto its homepage.

• Reduce Overall Costs
It will help to reduce a bank's costs in two fundamental ways: it minimize the cost of processing transactions and reduces the number of branches required to service an equivalent number of customer.

## 2.3.2 Safety Requirements

• Backup, recovery & business continuity
Banks should ensure adequate back up of data as may be required by their operations. Banks should also have, well documented and tested business continuity plans that address all aspects of the bank's business.

1. Both data and software should be backed up periodically, the frequency of back up depending on the recovery needs of the application. The back-up may be incremental or complete. Automating the backup procedures is preferred to obviate operator errors and missed back-ups.
2. Recovery and business continuity measures, based on criticality of the systems, should be in place and a documented plan with the organization and assignment of responsibilities of the key decision making personnel should exist.
3. An off-site back up is necessary for recovery from major failures / disasters to ensure business continuity. Depending on criticality, different technologies based on back up, hot sites, warm sites or cold sites should be available for business continuity. The business continuity plan should be frequently tested.

### 2.3.3 Security Requirements

We understand that there is nothing more important than knowing that transactions are private and secure. Therefore, we have applied the very latest in technology when creating the Internet Banking security architecture. The best way to understand the security architecture within the Internet Banking is to take it one step at a time. These security features are described briefly below.

- Account ID and Password (PIN) Protection

User Account ID and Password (PIN) protection occurs at the first level within the Internet Banking System. To access Internet Banking, users are required to enter an Account ID and password. Without these, access to the Internet Banking System is denied. Special password characters may be imposed by the Bank to provide a greater degree of security. The following characters may be used as required :!@#$%^&*()_+-=[]{}|\;:'",<.> /?
To further increase the level of security, the bank may impose a periodic change of passwords. If the Password Change option is imposed, a warning message will be displayed when logging-onto Internet Banking.

- Auto Timeout Screen Blanking

Although we recommend users never leave a PC unattended and financial information displayed while logged into Internet Banking, a built-in security feature minimizes the risk in such a situation. Users are required to acknowledge the message (Continue) presented in order to remain active in the Internet Banking session. The auto timeout feature warns users every 30 seconds prior to a pending timeout. If allowed to timeout, the Internet Banking session is halted and users are presented once again with the log-on screen.

- Sign-off Button

When an end-user is finished with Internet Banking, they should click the Sign-off button before going anywhere else on the Web. This ends the Internet Banking session.

- Failed Log-on Attempts

As an added security feature, the Internet Banking System is denied access after a pre-determined number of failed log-on attempts. If users have been locked out due to exceeding the pre-determined number of log-on attempts, the users must contact the Bank in order to be re-initialized.

- Encryption

In addition to password protection, we ensures server authentication by using the latest techniques of data encryption. Data encryption is a way of translating data into a form that is unintelligible without a deciphering mechanism.

## 3. CONCLUSION

In this paper, an analytical study for internet banking system has been presented. The banking system should be built within special requirements, since that the functional requirements and its specification has been proposed. The non-functional requirements represent the quality of the system but in internet banking system consider as most important requirements for the system. The security is one of these requirements which is considered as non-functional requirements and in many systems it's still not achieved. While in internet banking system it considered as one of the main requirements for the system what determine the success or fail of the system. Suggestion for best use for these requirements has been proposed in order to identify the requirements and how it is possible to achieve it.

## ACKNOWLEDGMENT

This research has been funded in part from multimedia University; the author would like to acknowledge the entire worker in this project, and the people who support in any way, the author would like to thanks his friends who has support in many ways

## REFERENCES

[1]  Bailey, J., & Pearson, S. (1983). Development of a Tool for Measuring and Analyzing Computer User Satisfaction. Management Science, 29(5), 530-545.

[2]  Beckett, A., Hewer, P., & Howcroft, B. (2000). An exposition of consumer behaviour in the financial services industry. The International Journal of Bank Marketing, 18(1).

[3]  Chin, J., Diehl, V., & Norman, L. (1988). Development of an instrument measuring user satisfaction of the human-computer interface. Paper presented at the Proceedings of the SIGCHI conference on Human factors in computing systems, New York.

[4]  Chin, W., & Lee, M. (2000). A proposed model and measurement instrument for the formation of IS satisfaction: the case of end-user computing satisfaction. Paper presented at the Proceedings of the twenty first international conference on Information systems, Brisbane, Australia.

[5]  Daniel, E. (1999). Provision of electronic banking in the UK and the Republic of Ireland. International Journal of Bank Marketing, 17(2), 72-82.

[6]  DeVellis, R. (2003). Scale Development: theory and applications (2 ed. Vol. 26). California: Sage Publications.

[7]  Dillman, D. (2000). Mail and Internet Surveys (2 ed.). New York: John Wiley & Sons.

[8]  Hadden, R., & Whalley, A. (2002). The Branch is dead, long live the Internet! (or so you'd have thought if we hadn't listened to the customer). International Journal of Market Research, 44(3), 283-297.

[9]  Hiltunen, M., Heng, L., & Helgesen, L. (2004). Personalized Electronic Banking Services. In C.-M. Karat, J. Blom & J. Karat (Eds.), Designing Personalized User Experiences in eCommerce (Vol. 5, pp. 119-140). Netherlands: Kluwer Academic Publishers.